\documentstyle[11pt]{article}

\advance \topmargin by -\headheight
\advance \topmargin by -\headsep
\evensidemargin \oddsidemargin
\begin{document}
\def\rf#1{(\ref{eq:#1})}
\def\lab#1{\label{eq:#1}}
\def\nonu{\nonumber}
\def\br{\begin{eqnarray}}
\def\er{\end{eqnarray}}
\def\be{\begin{equation}}
\def\ee{\end{equation}}
\def\eq{\!\!\!\! &=& \!\!\!\! }
\def\foot#1{\footnotemark\footnotetext{#1}}
\def\lb{\lbrack}
\def\rb{\rbrack}
\def\llangle{\left\langle}
\def\rrangle{\right\rangle}
\def\blangle{\Bigl\langle}
\def\brangle{\Bigr\rangle}
\def\llb{\left\lbrack}
\def\rrb{\right\rbrack}
\def\Blb{\Bigl\lbrack}
\def\Brb{\Bigr\rbrack}
\def\lcurl{\left\{}
\def\rcurl{\right\}}
\def\({\left(}
\def\){\right)}
\def\v{\vert}                     
\def\bv{\bigm\vert}               
\def\Bgv{\;\Bigg\vert}            
\def\bgv{\bigg\vert}              
\def\lskip{\vskip\baselineskip\vskip-\parskip\noindent}
\def\mskp{\par\vskip 0.3cm \par\noindent}
\def\sskp{\par\vskip 0.15cm \par\noindent}
\def\bc{\begin{center}}
\def\ec{\end{center}}
\def\Lbf#1{{\Large {\bf {#1}}}}
\def\lbf#1{{\large {\bf {#1}}}}

\def\tr{\mathop{\rm tr}}                  
\def\Tr{\mathop{\rm Tr}}                  
\newcommand\partder[2]{{{\partial {#1}}\over{\partial {#2}}}}
\newcommand\partderd[2]{{{\partial^2 {#1}}\over{{\partial {#2}}^2}}}
\newcommand\Bil[2]{\Bigl\langle {#1} \Bigg\vert {#2} \Bigr\rangle}  
\newcommand\bil[2]{\left\langle {#1} \bigg\vert {#2} \right\rangle} 
\newcommand\me[2]{\left\langle {#1}\right|\left. {#2} \right\rangle} 

\newcommand\sbr[2]{\left\lbrack\,{#1}\, ,\,{#2}\,\right\rbrack} 
\newcommand\Sbr[2]{\Bigl\lbrack\,{#1}\, ,\,{#2}\,\Bigr\rbrack} 
\newcommand\Gbr[2]{\Bigl\lbrack\,{#1}\, ,\,{#2}\,\Bigr\} } 
\newcommand\pbr[2]{\{\,{#1}\, ,\,{#2}\,\}}       
\newcommand\Pbr[2]{\Bigl\{ \,{#1}\, ,\,{#2}\,\Bigr\}}  
\newcommand\pbbr[2]{\lcurl\,{#1}\, ,\,{#2}\,\rcurl}  

\def\a{\alpha}
\def\b{\beta}
\def\c{\chi}
\def\d{\delta}
\def\D{\Delta}
\def\eps{\epsilon}
\def\vareps{\varepsilon}
\def\g{\gamma}
\def\G{\Gamma}
\def\grad{\nabla}
\def\h{{1\over 2}}
\def\k{\kappa}
\def\l{\lambda}
\def\L{\Lambda}
\def\m{\mu}
\def\n{\nu}
\def\o{\over}
\def\om{\omega}
\def\O{\Omega}
\def\p{\phi}
\def\vp{\varphi}
\def\P{\Phi}
\def\pa{\partial}
\def\tpa{{\tilde \partial}}
\def\bpa{{\bar \partial}}
\def\pr{\prime}
\def\ra{\rightarrow}
\def\lra{\longrightarrow}
\def\r{\rho}
\def\s{\sigma}
\def\S{\Sigma}
\def\t{\tau}
\def\th{\theta}
\def\bth{{\bar \theta}}
\def\Th{\Theta}
\def\z{\zeta}
\def\ti{\tilde}
\def\wti{\widetilde}
\def\adot{\stackrel{.}{\alpha}}    
\def\bdot{\stackrel{.}{\beta}} 
\def\gdot{\stackrel{.}{\gamma}}
\def\ddot{\stackrel{.}{\delta}} 
\newcommand\sumi[1]{\sum_{#1}^{\infty}}   
\newcommand\twomat[4]{\left(\begin{array}{cc}  
{#1} & {#2} \\ {#3} & {#4} \end{array} \right)}
\newcommand\threemat[9]{\left(\begin{array}{ccc}  
{#1} & {#2} & {#3}\\ {#4} & {#5} & {#6}\\
{#7} & {#8} & {#9} \end{array} \right)}
\newcommand\BDet[5]{\det_{{#1}}\left\Vert\begin{array}{cc}  
{#2} & {#3} \\ {#4} & {#5} \end{array} \right\Vert}   
\newcommand\Det[2]{\det_{{#1}} \left\Vert {#2} \right\Vert}
\newcommand\twocol[2]{\left(\begin{array}{cc}  
{#1} \\ {#2} \end{array} \right)}
\def\cA{{\cal A}}
\def\cB{{\cal B}}
\def\cC{{\cal C}}
\def\cD{{\cal D}}
\def\cE{{\cal E}}
\def\cF{{\cal F}}
\def\cG{{\cal G}}
\def\cH{{\cal H}}
\def\cI{{\cal I}}
\def\cJ{{\cal J}}
\def\cK{{\cal K}}
\def\cL{{\cal L}}
\def\cM{{\cal M}}
\def\cN{{\cal N}}
\def\cO{{\cal O}}
\def\cP{{\cal P}}
\def\cQ{{\cal Q}}
\def\cR{{\cal R}}
\def\cS{{\cal S}}
\def\cT{{\cal T}}
\def\cU{{\cal U}}
\def\cV{{\cal V}}
\def\cX{{\cal X}}
\def\cW{{\cal W}}
\def\cY{{\cal Y}}
\def\cZ{{\cal Z}}

\def\mark{\noindent{\bf Remark.}\quad}
\def\prop{\noindent{\bf Proposition.}\quad}
\def\exam{\noindent{\bf Example.}\quad}
\newtheorem{definition}{Definition}[section]
\newtheorem{proposition}{Proposition}[section]
\newtheorem{theorem}{Theorem}[section]
\newtheorem{lemma}{Lemma}[section]
\newtheorem{corollary}{Corollary}[section]
\def\proof{\par{\it Proof}. \ignorespaces} \def\endproof{{$\Box$}\par}
\newenvironment{Proof}{\proof}{\endproof} 

\def\symp#1{{\cal S}{\cal D}if\!\! f \, ({#1})}
\def\esymp#1{{\wti {\cal S}{\cal D}if\!\! f} \, ({#1})}
\def\Symp#1{{\rm SDiff}\, ({#1})}
\def\eSymp#1{{\wti {\rm SDiff}}\, ({#1})}
\def\vol#1{{{\cal D}if\!\! f}_0 ({#1})}
\def\Vol#1{{\rm Diff}_0 ({#1})}

\def\Winf{{\bf W_\infty}}               
\def\Win1{{\bf W_{1+\infty}}}           
\def\nWinf{{\bf {\hat W}_\infty}}       
\def\PsDA{\Psi{\cal DO}}
\def\bD{{\bar D}}

\newcommand{\nit}{\noindent}
\newcommand{\ct}[1]{\cite{#1}}
\newcommand{\bi}[1]{\bibitem{#1}}
\newcommand\PRL[3]{{\sl Phys. Rev. Lett.} {\bf#1} (#2) #3}
\newcommand\NPB[3]{{\sl Nucl. Phys.} {\bf B#1} (#2) #3}
\newcommand\NPBFS[4]{{\sl Nucl. Phys.} {\bf B#2} [FS#1] (#3) #4}
\newcommand\CMP[3]{{\sl Commun. Math. Phys.} {\bf #1} (#2) #3}
\newcommand\PRD[3]{{\sl Phys. Rev.} {\bf D#1} (#2) #3}
\newcommand\PLA[3]{{\sl Phys. Lett.} {\bf #1A} (#2) #3}
\newcommand\PLB[3]{{\sl Phys. Lett.} {\bf #1B} (#2) #3}
\newcommand\JMP[3]{{\sl J. Math. Phys.} {\bf #1} (#2) #3}
\newcommand\PTP[3]{{\sl Prog. Theor. Phys.} {\bf #1} (#2) #3}
\newcommand\SPTP[3]{{\sl Suppl. Prog. Theor. Phys.} {\bf #1} (#2) #3}
\newcommand\AoP[3]{{\sl Ann. of Phys.} {\bf #1} (#2) #3}
\newcommand\RMP[3]{{\sl Rev. Mod. Phys.} {\bf #1} (#2) #3}
\newcommand\PR[3]{{\sl Phys. Reports} {\bf #1} (#2) #3}
\newcommand\FAP[3]{{\sl Funkt. Anal. Prilozheniya} {\bf #1} (#2) #3}
\newcommand\FAaIA[3]{{\sl Functional Analysis and Its Application} {\bf #1}
(#2) #3}
\def\TAMS#1#2#3{{\sl Trans. Am. Math. Soc.} {\bf #1} (#2) #3}
\def\InvM#1#2#3{{\sl Invent. Math.} {\bf #1} (#2) #3}
\def\AdM#1#2#3{{\sl Advances in Math.} {\bf #1} (#2) #3}
\def\PNAS#1#2#3{{\sl Proc. Natl. Acad. Sci. USA} {\bf #1} (#2) #3}
\newcommand\LMP[3]{{\sl Letters in Math. Phys.} {\bf #1} (#2) #3}
\newcommand\IJMPA[3]{{\sl Int. J. Mod. Phys.} {\bf A#1} (#2) #3}
\newcommand\TMP[3]{{\sl Theor. Mat. Phys.} {\bf #1} (#2) #3}
\newcommand\JPA[3]{{\sl J. Physics} {\bf A#1} (#2) #3}
\newcommand\JSM[3]{{\sl J. Soviet Math.} {\bf #1} (#2) #3}
\newcommand\MPLA[3]{{\sl Mod. Phys. Lett.} {\bf A#1} (#2) #3}
\newcommand\JETP[3]{{\sl Sov. Phys. JETP} {\bf #1} (#2) #3}
\newcommand\JETPL[3]{{\sl  Sov. Phys. JETP Lett.} {\bf #1} (#2) #3}
\newcommand\PHSA[3]{{\sl Physica} {\bf A#1} (#2) #3}
\newcommand\PHSD[3]{{\sl Physica} {\bf D#1} (#2) #3}
\newcommand\JPSJ[3]{{\sl J. Phys. Soc. Jpn.} {\bf #1} (#2) #3}
\newcommand\JGP[3]{{\sl J. Geom. Phys.} {\bf #1} (#2) #3}

\pagestyle{empty}

\vspace*{-1cm}
\noindent
March, 1999 \hfill{BGU-99 / 10 / Mar - PH}\\
\phantom{bla}
\hfill{hep-th/9903245}
\\
\begin{center}
{\large {\bf Composite Vector and Tensor Gauge Fields, and Volume-Preserving 
Diffeomorphisms}}  
\end{center}    
 
\begin{center}
E.I. Guendelman${}^1$, E. Nissimov${}^{1,2}$ and S. Pacheva${}^{1,2}$ \\    
{\em ${}^1$Department of Physics, Ben-Gurion University of the Negev }\\
{\em Box 653, IL-84105 $\;$Beer Sheva, Israel} \\
{\em E-Mail}: guendel@bgumail.bgu.ac.il , emil@bgumail.bgu.ac.il ,
svetlana@bgumail.bgu.ac.il \\
{\em ${}^2$ Institute of Nuclear Research and Nuclear Energy} \\
{\em Boul. Tsarigradsko Chausee 72, BG-1784 $\;$Sofia, Bulgaria} \\
{\em E-Mail}: nissimov@inrne.bas.bg , svetlana@inrne.bas.bg
\end{center} 


\begin{abstract}
We describe new theories of composite vector and tensor ({\em p-form}) gauge 
fields made out of zero-dimensional constituent scalar fields (``primitives'').
The local gauge symmetry is replaced by an {\em infinite-dimensional global 
Noether symmetry} -- the group of volume-preserving (symplectic) 
diffeomorphisms of the target space of the scalar primitives.
We find additional non-Maxwell and non-Kalb-Ramond solutions describing 
topologically massive tensor gauge field configurations in odd space-time
dimensions. Generalization to the supersymmetric case is also sketched.
\end{abstract}

\underline{{\bf 1. Introduction}} ~~Infinite-dimensional symmetries play an
increasingly important r\^{o}le in various areas of physics. The current 
interest towards them is motivated mainly due to the recent discovery of
effective description of Seiberg-Witten theory in terms of integrable models 
in lower space-time dimensions \ct{SW-integr}, as well as the r\^{o}le of 
$\Win1$-algebra in the field-theoretic description of the quantum Hall effect 
\ct{W-QHE}.

The hallmark of the completely integrable two-dimensional field-theoretic 
models is the presence of infinite sets of conservation laws allowing for
their exact solvability. It is however difficult to find a realistic field 
theory in $D\! =\! 4$ space-time dimensions possessing an infinite number of 
nontrivial conserved charges.
In the present note (which is an extension of our work in ref.\ct{area})
we will construct a series of new theories in higher space-time dimensions
allowing for an infinite number of conservation laws, which at the same time
closely resemble the standard vector and tensor ({\em p-form}) gauge theories.
The basic idea is to introduce the ordinary gauge fields as composite fields
built up of more elementary ``primitive'' scalar field constituents and 
to replace the ordinary local gauge symmetry with a {\em global 
infinite-dimensional} Noether symmetry acting on the target space of the 
scalar `primitives''.

Let us briefly recall some basic notions connected with the 
infinite-dimensional groups $\Vol{\cT^{s}}$ of volume-preserving 
diffeomorphisms on ($s$-dimensional) smooth manifolds $\cT^s$.
$\Vol{\cT^{s}}$ is defined as the group of all diffeomorphisms 
preserving the canonical volume form 
$\frac{1}{s!}\vareps_{a_1 \ldots a_s} d\vp^{a_1} \wedge\ldots\wedge d\vp^{a_s}$
(with $\lcurl \vp^a \rcurl_{a=1}^s$ being a set of local coordinates) :
\be
\Vol{\cT^{s}} \equiv \lcurl  \vp^a \to G^a (\vp ) \; ;\;
\vareps_{b_1 \ldots b_s} \partder{G^{b_1}}{\vp^{a_1}} \cdots
\partder{G^{b_s}}{\vp^{a_s}} = \vareps_{a_1 \ldots a_s}  \rcurl
\lab{vol-diff-group}
\ee
Accordingly, the Lie algebra $\vol{\cT^{s}}$ of infinitesimal volume-preserving
diffeomorphisms is given by:
\be
\vol{\cT^{s}} \equiv \lcurl \G^a (\vp ) \;\; ;\;\;
G^a (\vp ) \approx \vp^a + \G^a (\vp ) \; ,\; \partder{\G^b}{\vp^b} =0 \rcurl
\lab{vol-diff-alg} 
\ee
{\sl i.e.}, 
\be
\G^a (\vp ) = \frac{1}{(s-2)!} \vareps^{a b c_1 \ldots c_{s-2}}
\partder{}{\vp^b}\G_{c_1 \ldots c_{s-2}}(\vp )
\lab{rot}
\ee
In the simplest case $s=2$ the algebra $\vol{\cT^{2}}$ coincides with the
algebra of symplectic (area-preserving) diffeomorphisms~ 
$\symp{\cT^{2}} \equiv \lcurl \G (\vp ) \; ;\; \sbr{\G_1}{\G_2} \equiv
\pbbr{\G_1}{\G_2}=\vareps^{ab}\partder{\G_1}{\vp^a}\partder{\G_2}{\vp^b}\rcurl$
which contains as a subalgebra the centerless conformal Virasoro algebra and
whose Lie-algebraic deformation is the well-known $\Win1$-algebra.

\underline{{\bf 2. Composite p-form Gauge Theories}}~~
Let us consider a set of $s(\equiv p+1)$ zero-dimensional scalar fields
$\lcurl \vp^a (x) \rcurl_{a=1}^s$ taking values in a smooth manifold
$\cT^s$. The pull-back of its canonical volume $s(\equiv p+1)$-form to
Minkowski space-time gives rise to an antisymmetric $s$-tensor gauge field 
strength and its associated antisymmetric $(s-1)$-tensor gauge potential:
\br
\frac{1}{s!} \vareps_{a_1 \ldots a_s} d\vp^{a_1} \wedge\cdots\wedge d\vp^{a_s}
= \frac{1}{s!} F_{\m_1\ldots \m_s}(\vp ) dx^{\m_1} \wedge\cdots\wedge dx^{\m_s}
\lab{s-diff-form} \\
F_{\m_1\ldots \m_s} (\vp ) = \vareps_{a_1 \ldots a_s}\,
\pa_{\m_1} \vp^{a_1}\ldots \pa_{\m_s} \vp^{a_s}
\lab{F-primitiv-s} \\
F_{\m_1\ldots \m_s} (\vp ) = s \pa_{\lb \m_1} A_{\m_2\ldots \m_s\rb} (\vp )
\quad ,\quad  A_ {\m_1\ldots \m_{s-1}} (\vp )=
{1\over s} \vareps_{a_1 \ldots a_s} \vp^{a_1} \pa_{\m_1} \vp^{a_2}\ldots
\pa_{\m_{s-1}} \vp^{a_s}
\lab{A-primitiv-s}
\er
where the square brackets indicate total antisymmetrization of indices.
One can easily verify that the field strength \rf{F-primitiv-s} is invariant 
under arbitrary field transformations (reparametrizations) 
$\vp^a (x) \to G^a \(\vp (x)\)$ belonging to the infinite-dimensional group 
$\Vol{\cT^s}$ \rf{vol-diff-group}, whereas its potential \rf{A-primitiv-s} 
undergoes a $\vp$-dependent $(s-2)$-rank local gauge transformation 
(cf. \rf{vol-diff-alg}--\rf{rot}):
\be
A_ {\m_1\ldots \m_{s-1}} (\vp )\to A_ {\m_1\ldots \m_{s-1}} (\vp )+
(s-1)^{-1} \pa_{\lb \m_1} \L_{\m_2\ldots \m_{s-1}\rb} (\vp )
\lab{transform-s} 
\ee
\br
\L_{\m_1\ldots \m_{s-2}} (\vp )=
\llb \( 1 - {1\over s} \vp^b \partder{}{\vp^b}\) \G_{a_1\ldots a_{s-2}}(\vp )
- \frac{(s-1)(s-2)}{s} \vp^b \partder{}{\vp^{\lb a_1}}
\G_{\v b\v a_2\ldots a_{s-2}\rb}\rrb \times  \nonu \\
\times \,\, \pa_{\m_1}\vp^{a_1}\ldots \pa_{\m_{s-2}}\vp^{a_{s-2}}
\lab{gauge-s}
\er
In the simplest case $s\! =\! 2$ the composite electromagnetic field 
strenght and potential read:
\br
F_{\m\n} (\vp )= \vareps_{ab} \pa_\m \vp^a \pa_\n \vp^b   \qquad ,\qquad
F_{\m\n} (\vp )= \pa_\m A_\n (\vp )- \pa_\n A_\m (\vp )
\lab{F-primitiv} \\
A_\m (\vp )= \h \vareps_{ab} \vp^a \pa_\m \vp^b      \qquad, \qquad
A_\m (\vp ) \to  A_\m (\vp ) + \pa_\m \( \G (\vp ) - 
\h \vp^a \partder{\G}{\vp^a}\)
\lab{A-primitiv}
\er

Now it is straightforward, upon using \rf{F-primitiv-s}--\rf{A-primitiv-s},
to construct field-theory models of arbitrary
{\em p-form} tensor gauge fields involving the scalar ``primitives'' $\vp^a$
coupled to ordinary matter ({\sl e.g.}, fermionic) fields, where the
standard local {\em p-form} tensor gauge invariance is substituted with
the {\em global infinite-dimensional} Noether symmetry of
volume-preserving diffeomorphisms on the $s\!\equiv\! p+1$-dimensional target
space of primitive scalar constituents. The simplest model (for $s\! =\! 2$)
is the so called ``mini-QED'' \ct{area} :
$\cL = -{1\over {4e^2}} F_{\m\n}^2 (\vp ) + {\bar \psi} \( i{\not\!\!\pa} -
{\not\!\! A}(\vp ) -im\) \psi$ ,
where $F_{\m\n}(\vp )$ and $ A_\m (\vp )$ are given by \rf{F-primitiv} and
\rf{A-primitiv}, respectively. Similarly, the arbitrary higher rank 
{\em p-form} composite tensor gauge field theories are defined by:
\be
S = -{1\o{2(p+1) e^2}} \int d^D x\, F^2_{\m_1\ldots\m_{p+1}} (\vp (x)) +
\int d^D x\, A_{\m_1\ldots\m_p}(\vp (x))\, J^{\m_1\ldots\m_p}_{matter}(x) +
S_{matter}
\lab{comp-p-form}
\ee
where $F_{\m_1\ldots\m_{p+1}} (\vp (x))$ and $A_{\m_1\ldots\m_p}(\vp (x))$
are given by \rf{F-primitiv-s}--\rf{A-primitiv-s}. In particular, for 
$s\! =\! 3$
we have a ``mini-Kalb-Ramond'' model:
\be
\cS = - \frac{1}{3!e^2} \int d^D x \, F^2_{\m\n\l} (\vp (x))  +
\int d^2 \s \, {1\over 3} \vareps_{abc}\, \vp^a (x(\s ))
\pa_{\s_1} \vp^b (x(\s )) \pa_{\s_2} \vp^c (x(\s ))
\lab{mini-KR}
\ee
where the second integral is over the string world-sheet given by
$x^\m = x^\m (\s )$. 

Let us particularly stress that, although the Lagrangians of the composite
{\em p-form} gauge theories \rf{comp-p-form}--\rf{mini-KR} contain higher 
order derivatives w.r.t. $\vp^a$, they are only quadratic w.r.t. 
time-derivatives.
Also, note that the Chern-Simmons terms and ``topolocal'' densities for the
composite {\em p-form} tensor gauge fields \rf{F-primitiv-s}--\rf{A-primitiv-s}
{\em identically} vanish in space-time dimensions $D\! =\! 2p+1$ and 
$D\! =\! 2p+2$, respectively, {\sl e.g.} :
\br
\vareps^{\m_1 \ldots \m_{2p+1}} A_{\m_1\ldots\m_p}(\vp )
F_{\m_{p+1}\ldots\m_{2p+1}}(\vp ) =    \nonu  \\
{1\o{p+1}} \vareps_{a_1\ldots a_{p+1}} \,\vareps_{b_1\ldots b_{p+1}}\,
\vareps^{\m_1 \ldots \m_{2p+1}} \vp^{a_1} \pa_{\m_1}\vp^{[a_2}\ldots 
\pa_{\m_p}\vp^{a_{p+1}} \pa_{\m_{p+1}}\vp^{b_1} \ldots
\pa_{\m_{2p+1}}\vp^{b_{p+1}]} = 0 
\lab{CS-vanish}
\er
due to the total antisymmetrization of the $2p+1$ indices
$a_2,\ldots ,a_{p+1},b_1,\ldots ,b_{p+1}$ taking only $p+1$ values.

\underline{{\bf 3. Non-Maxwell and Non-Kalb-Ramond Solutions}}~~
The action \rf{comp-p-form} yields the following equations of motion:
$\vareps_{a b_1\ldots b_p} \pa_{\m_1}\vp^{b_1}\ldots \pa_{\m_p}\vp^{b_p}
\llb {1\o{e^2}} \pa_\n F^{\n \m_1 \ldots \m_p}(\vp (x)) +
J^{\m_1\ldots\m_p}_{matter}(x) \rrb = 0$ 
upon variation w.r.t. $\vp^a$ or, equivalently, upon multiplying the l.h.s. of
the last equation by $\vp^a$ and accouting for \rf{A-primitiv-s} :
\be
A_{\m_1\ldots\m_p}(\vp (x)) 
\llb {1\o{e^2}} \pa_\n F^{\n \m_1 \ldots \m_p}(\vp (x)) +
J^{\m_1\ldots\m_p}_{matter}(x) \rrb = 0
\lab{EM-p-form}
\ee
From \rf{EM-p-form} we notice that any solution $A_{\m_1\ldots\m_p}(x)$
of the standard {\em p-form} gauge theory, {\sl i.e.}, such 
$A_{\m_1\ldots\m_p}(x)$ for which the term in the square brackets in 
\rf{EM-p-form} vanishes, is  authomatically a solution of the new composite 
{\em p-form} gauge theory \rf{comp-p-form} provided $A_{\m_1\ldots\m_p}(x)$ is
representable (up to $(p-1)$-rank gauge transformation) in terms of scalar 
``primitives'' as in \rf{A-primitiv-s}. 

On the other hand, Eqs.\rf{EM-p-form} possess
additional solutions unattainable in ordinary {\em p-form} gauge theories,
namely such $A_{\m_1\ldots\m_p}(\vp (x))$ for which the corresponding 
factor in the square brackets in \rf{EM-p-form} is non-zero:
\be
{1\o{e^2}} \pa_\n F^{\n \m_1 \ldots \m_p}(\vp (x)) +
J^{\m_1\ldots\m_p}_{matter}(x) + \cJ^{\m_1\ldots\m_p}(\vp (x)) = 0
\lab{EM-p-form-1}
\ee
where the additional ``current'' $\cJ^{\m_1\ldots\m_p}$ obeys:
$A_{\m_1\ldots\m_p}(\vp (x))\, \cJ^{\m_1\ldots\m_p}(\vp (x)) = 0$.
An immediate nontrivial example for $\cJ^{\m_1\ldots\m_p}$ satisfying 
the latter constraint in odd space-time dimensions $D=2p+1$ is
(due to \rf{CS-vanish}) :
\be
\cJ^{\m_1\ldots\m_p}(\vp (x)) = {1\o{(p+1)!}} 
\vareps^{\m_1\ldots\m_p \n_1\ldots\n_{p+1}} F_{\n_1\ldots\n_{p+1}}(\vp (x))
\lab{CS-current}
\ee
so that Eqs.\rf{EM-p-form-1} become the eqs. of motion of ordinary {\em p-form}
gauge theory with {\em additional} Chern-Simmons term. Thus, if we succeed to 
find a solution to:
\be
{1\o{e^2}} \pa_\n F^{\n \m_1 \ldots \m_p} +
{1\o{(p+1)!}}\vareps^{\m_1\ldots\m_p \n_1\ldots\n_{p+1}} F_{\n_1\ldots\n_{p+1}}
+ J^{\m_1\ldots\m_p}_{matter} = 0
\lab{EM-p-form-CS}
\ee
which is expressible in terms of scalar ``primitives'', we therefore obtain
a non-Maxwell/non-Kalb-Ramond solution of Eqs.\rf{EM-p-form}.

We will present here such a solution of Eqs.\rf{EM-p-form} in the free case
$J^{\m_1\ldots\m_p}_{matter}(x) = 0$ (for simplicity, we consider the 
``mini-Kalb-Ramond'' model \rf{mini-KR}) :
\be
A^{\m\n}(\vp )\, \pa^\l F_{\l\m\n}(\vp ) = 0 \; ,\;\;
\pa^\l F_{\l\m\n}(\vp) + {{e^2}\o{3!}}\vareps_{\m\n\r\s\t} F^{\r\s\t}(\vp) = 0
\; ,\;\; 
F_{\l\m\n}(\vp) = \vareps_{abc}\pa_\l \vp^a \pa_\m \vp^b \pa_\n \vp^c \nonu\\
\phantom{aaaaaaa}
\lab{EM-2-form}
\ee
\be
\vp^1 (x) = k_\m x^\m \quad , \quad \vp^2 (x) = \ell_\m x^\m \quad , \quad
\vp^3 (x) = x_\m \llb p^\m \sin (k.x) + q^\m \cos (k.x) \rrb
\lab{vp-sol} 
\ee
where $k,\ell ,p,q$ are constant 5-vectors obeying:
\br
p.k = p.\ell = q.k = q.\ell = k.\ell = p.q = \ell^2 = 0   \quad ,\quad
p^2 = q^2   \nonu  \\
k^2 \( p_\m \ell_\n - p_\n \ell_\m\) - 
e^2 \vareps_{\m\n\r\s\t} k^{\r}\ell^{\s}q^{\t} = 0   \quad ,\quad
k^2 \( q_\m \ell_\n - q_\n \ell_\m\) + 
e^2 \vareps_{\m\n\r\s\t} k^{\r}\ell^{\s}p^{\t} = 0 
\lab{const-constraints}
\er
Hence, we find from Eqs.\rf{EM-2-form}--\rf{vp-sol} that composite 
{\em p-form} tensor gauge theories \rf{comp-p-form} exhibit dynamical parity 
breakdown since they carry ``topologically'' massive excitations.

\underline{{\bf 4. Supersymmetric Generalization}}~~
Here we will briefly indicate the supersymmetric generalization of
\rf{comp-p-form} in the simplest case of interest: $s\! =\! 2$ 
(``mini-super-QED'') and $D\! =\! 4$ (for superspace notations and 
definitions, see {\sl e.g.} \ct{Wess-Bagger}). The superfield of the vector 
supermultiplet $\bigl( A_\m ,\l_\a, {\bar \l}^{\adot}\bigr)$ :
\br
V(x,\th,\bth ) = a(x) + \th_\a \chi^\a (x) + 
\bth_{\adot} {\bar\chi}^{\adot}(x) + \d (\th) {\bar b}(x) - \d (\bth) b(x) 
+ (\th \s^\m \bth) A_\m (x) + \nonu  \\
+\, \d (\bth) \th_\a \bigl(\l^\a (x) + i\pa^{\a\bdot}{\bar \chi}_{\bdot}(x)\bigr)
- \d (\th) \bth_{\adot} \bigl({\bar \l}^{\adot} (x) + 
i\pa^{\adot\b}\chi_{\b}(x)\bigr) + 
\d (\th) \d (\bth) \bigl(\cD (x) - \pa^2 a(x)\bigr)
\lab{V-SF}
\er
(here $a,b,{\bar b},\chi^\a,{\bar\chi}^{\adot}$ and $\cD$
are pure-gauge and auxiliary component fields, respectively, and
$\d (\th) \equiv \h \th_\a \th^\a\, ,\,\d (\bth)$ 
$\equiv \h \bth^{\adot} \,\bth_{\adot}$)
is constructed as bilinear composite in terms of a pair of chiral and
anti-chiral ``primitive'' scalar superfields:
\br
\P (x,\th,\bth ) = e^{-i(\th \s^\m \bth)\pa_\m}
\llb \vp (x) + \th_\a \psi^\a + \d (\th) f(x) \rrb
\lab{chiral-SF} \\
{\bar \P} (x,\th,\bth ) = e^{i(\th \s^\m \bth)\pa_\m}
\llb {\bar \vp} (x) + \bth_{\adot} {\bar \psi}^{\adot}
- \d (\bth) {\bar f}(x) \rrb
\lab{antichiral-SF}
\er
through the following simple formula:
\be
V(\P,{\bar\P}) = \P\, {\bar\P}
\lab{SF-primitiv}
\ee
In terms of component fields Eq.\rf{SF-primitiv} reads:
\be
A_\m = -i {\bar\vp} {\stackrel{\leftrightarrow}{\pa}}_\m \vp +
{\bar \psi}^{\adot} (\s_\m)_{\a\adot} \psi^\a  \quad ,\quad
\l^\a = -2i {\bar \psi}_{\bdot} \pa^{\bdot\a} \vp - f \psi^\a 
\quad ,\quad
{\bar \l}^{\adot} = -2i \psi_\b \pa^{\adot\b} {\bar \vp} - 
{\bar f} {\bar \psi}^{\adot}
\lab{A-lambda-primitiv}
\ee
which are the supersymmetric generalization of \rf{A-primitiv}. The superspace
action of ``mini-super-QED'' is obtained by replacing the superfield $V$
\rf{V-SF} with its composite form \rf{SF-primitiv} in the well-known
super-QED action:
\be
S = - {1\o{16e^2}} 
\int d^4 x\, d^2 \th\, W^\a (\P,{\bar \P})\, W_\a (\P,{\bar \P})
+ \int d^4 x\, d^2 \th\, d^2 \bth \,{\bar \cM}\, e^{V (\P,{\bar \P})}\,\cM
\lab{mini-super-QED}
\ee
where $W_\a \! =\! \h \bD^{\bdot} \bD_{\bdot} D_\a V$ is the supersymmetric
field strenght (with
$D^\a \! =\! \partder{}{\th_\a} - i \pa^{\a\bdot}\bth_{\bdot}$ and
$\bD_{\adot}\! =\! \partder{}{\bth^{\adot}} - i \pa_{\b\adot}\th^{\b}$ 
being the standard super-derivatives) and $\cM,{\bar \cM}$ are the
chiral/antichiral matter superfields.

\underline{{\bf 5. Outlook}}~~ It is an interesting task to study in detail
the supersymmetric composite {\em p-form} gauge theories generalizing 
\rf{mini-super-QED}. Of particular interest to modern string theory would be
to find supersymmetric $p$-brane solutions in these theories extending the
work in ref.\ct{Castro}, where $p^\pr$-brane solutions have been obtained in
the purely bosonic composite {\em p-form} tensor gauge theories
\rf{comp-p-form} when $D\! = \! (p+1) + (p^\pr +1)$.
On the other hand, the possibility of defining volume
forms in terms of ``primitive'' scalar fields as in \rf{F-primitiv}
suggests the idea of replacement of the standard measure of integration
$\sqrt{-g}$ in general-coordinate invariant theories by the
``primitive'' scalar composite $\det \Vert \pa_\m \vp^a \Vert$ (here 
$p+1\! =\! D$). The basic reasons and advantages of this approach are
discussed in separate contributions to these Proceedings (by E.I. Guendelman
and A.B. Kaganovich and by E.I.Guendelman)). 

\end{document}